\documentclass[prd,aps,eqsecnum,floatfix,nofootinbib,preprint,tightenlines]{revtex4}

\usepackage{latexsym}
\usepackage{graphicx}
\usepackage{multirow}
\usepackage{amsmath}

\def\bibi{\bibitem}
\def\floatcaption#1#2{ \caption{#2 \label{#1}} }

\let\inodot=\i

\def\a{\alpha}
\def\b{\beta}
\def\c{\chi}
\def\d{\delta}
\def\g{\gamma}

\def\i{\iota}

\def\m{\mu}
\def\n{\nu}

\def\p{\pi}                     % Also, \varpi
\def\r{\rho}                    %       \varrho
\def\s{\sigma}                  %       \varsigma
\def\t{\tau}

\def\D{\Delta}

\def\P{\Pi}

\def\cq{{\cal Q}}

\def\cbo{{\,\raise-.15ex\Sc [\,}}                       % curly "
\def\ddt#1{{\buildrel {\hbox{\LARGE .\kern-2pt.}} \over {#1}}}% double dot-over
\def\ie{\mbox{\it i.e.}}

\def\etc{\mbox{\it etc.}}

\def\half{{1\over 2}}

\long\def\symbolfootnote[#1]#2{\begingroup%
\def\thefootnote{\fnsymbol{footnote}}\footnote[#1]{#2}\endgroup}

\long \def \blockcomment #1\endcomment{}

\def\seef{{\it cf.}}
\def\tP{{\tilde{\P}}}

\def\ta{{\tilde{a}}}

\begin{document}

\thispagestyle{empty}

\begin{center}
\vspace*{5mm}
\begin{boldmath}
{\large\bf Tests of hadronic vacuum polarization fits for the muon anomalous
magnetic moment}
\end{boldmath}
\\[10mm]
Maarten Golterman,$^{a,b}$
Kim Maltman,$^{c,d}$ Santiago Peris$^e$
\\[8mm]
{\small\it
$^a$Institut de F\'\inodot sica d'Altes Energies (IFAE), 
Universitat Aut\`onoma de Barcelona\\ E-08193 Bellaterra, Barcelona, Spain
\\[5mm]
$^b$Department of Physics and Astronomy,
San Francisco State University\\ San Francisco, CA 94132, USA
\\[5mm]
$^c$Department of Mathematics and Statistics,
York University\\  Toronto, ON Canada M3J~1P3
\\[5mm]
$^d$CSSM, University of Adelaide, Adelaide, SA~5005 Australia
\\[5mm]
$^e$Department of Physics, Universitat Aut\`onoma de Barcelona\\ E-08193 Bellaterra, Barcelona, Spain}
\\[10mm]
{ABSTRACT}
\\[2mm]
\end{center}

\begin{quotation}
Using experimental spectral data for hadronic $\t$ decays from the OPAL
experiment, supplemented by a  phenomenologically 
successful parameterization for the high-$s$ region not covered by 
the data, we construct a physically constrained model of the isospin-one 
vector-channel polarization function.  Having such
a model as a function of Euclidean momentum 
$Q^2$ allows us to explore the systematic error associated with  
fits to the $Q^2$ dependence of lattice data for the hadronic
electromagnetic current polarization function which have been used
in attempts to compute the leading order hadronic contribution, 
$a_\m^{\rm HLO}$, to the muon anomalous magnetic moment.
In contrast to recent claims made in the literature, we find that a 
final error in this quantity of the order of a few percent does not 
appear possible with current lattice data, given the present lack of 
precision in the determination of the vacuum polarization at low $Q^2$.
We also find that fits to the vacuum polarization using fit functions based on
Vector Meson Dominance are unreliable, in that the fit error on 
$a_\m^{\rm HLO}$ is typically much smaller than the difference 
between the value obtained from the fit and the exact model value. 
The use of a sequence of
Pad\'e approximants known to converge 
to the true vacuum polarization  
appears to represent a more promising approach. 

\end{quotation}

%%####%%
\newpage
\section{\label{intro} Introduction}
%%####%%
In the quest for a precision computation of the muon anomalous magnetic moment $a_\m=(g-2)/2$, the contribution from the hadronic vacuum
polarization at lowest order in the fine-structure constant $\a$,
$a_\m^{\rm HLO}$, plays
an important role.   While the contribution itself is rather small (of order
0.06 per mille!) the error in this contribution dominates the total
uncertainty in the present estimate of the Standard-Model value.
In order to reduce this uncertainty, and resolve or solidify  the potential discrepancy
between the experimental and Standard-Model values, it is thus important
to corroborate, and if possible improve on, the total error in $a_\m^{\rm HLO}$.

Recently, there has been much interest in computing this quantity using 
Lattice QCD \cite{TB12}.   In terms of the vacuum polarization $\P^{\rm em}(Q^2)$ at Euclidean momenta $Q^2$, $a_\m^{\rm HLO}$ is given by
the integral \cite{TB2003,ER}
\begin{eqnarray}
\label{amu}
a_\m^{\rm HLO}&=&4\a^2\int_0^\infty dQ^2\,f(Q^2)\left(\P^{\rm em}(0)-\P^{\rm em}(Q^2)\right)\ ,\\
f(Q^2)&=&m_\m^2 Q^2 Z^3(Q^2)\,\frac{1-Q^2 Z(Q^2)}{1+m_\m^2 Q^2 Z^2(Q^2)}\ ,\nonumber\\
Z(Q^2)&=&\left(\sqrt{(Q^2)^2+4m_\m^2 Q^2}-Q^2\right)/(2m_\m^2 Q^2)\ ,\nonumber
\end{eqnarray}
where $m_\m$ is the muon mass, and for non-zero momenta
$\P^{\rm em}(Q^2)$ is defined from the hadronic contribution to the
electromagnetic vacuum polarization $\P^{\rm em}_{\m\n}(Q)$,
\begin{equation}
\label{Pem}
\P^{\rm em}_{\m\n}(Q)=\left(Q^2\d_{\m\n}-Q_\m Q_\n\right)\P^{\rm em}(Q^2)
\end{equation}
in momentum space.

Since the integral is over Euclidean momentum, this is an ideal task for
the lattice, if $\P^{\rm em}(Q^2)$ can be computed at sufficiently many
non-zero values of $Q^2$, especially in the region $Q^2\sim m^2_\m$ 
which dominates the integral. However, because of the necessity 
of working in a finite volume, momenta are quantized on the lattice, 
which turns out to make this a difficult problem. Figure~\ref{f1} 
demonstrates the problem. On the left, we see a typical form of the 
subtracted vacuum polarization, together with the low-$Q^2$ points 
from a typical lattice data set.\footnote{For
the curve and data shown here, see Sec.~\ref{model} and Sec.~\ref{lattice}.}
On the right, we see the same information, but now multiplied by the
weight $f(Q^2)$ in Eq.~(\ref{amu}).

%%%%%%%%%%%%%%%%%%%
\begin{figure}[t]
\centering
\includegraphics[width=2.9in]{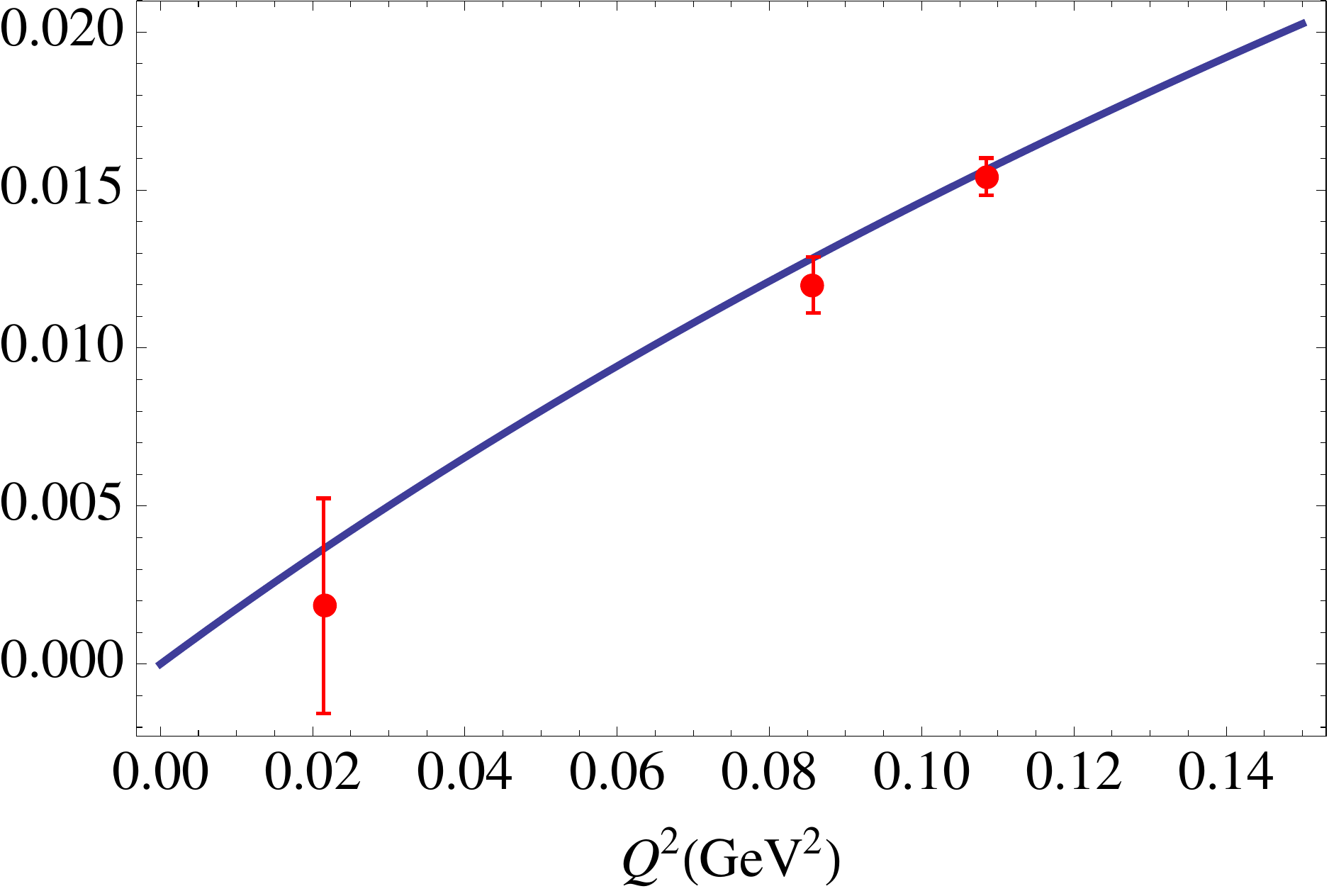}
\hspace{.1cm}
\includegraphics[width=2.9in]{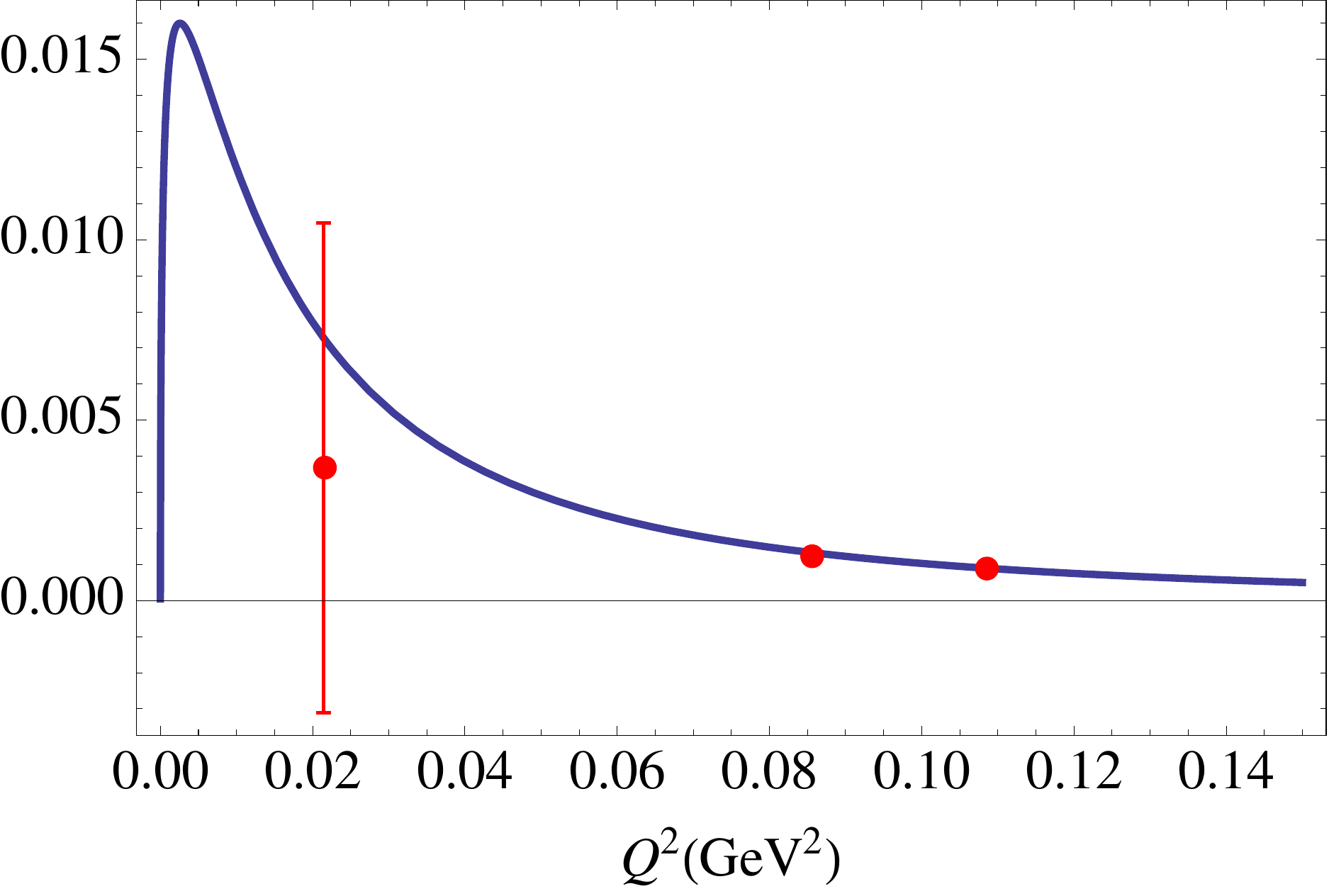}
\floatcaption{f1}{Low-$Q^2$ behavior of the subtracted vacuum
polarization $\P(0)-\P(Q^2)$ (left panel) and of the integrand 
$ f(Q^2)(\P(0)-\P(Q^2))$ in Eq.~(\ref{amu})
(right panel).  Red points show typical data on a $64^3\times 144$
lattice with lattice spacing $0.06$~fm and periodic boundary conditions.}
\vspace*{2ex}
\end{figure}
%%%%%%%%%%%%%%%%%%%

Figure~\ref{f1} clearly shows why evaluating the integral in Eq.~(\ref{amu})
as a Riemann sum using typical lattice data is ruled out.   In principle,
going to larger volumes, or 
using twisted boundary conditions  \cite{DJJW2012,ABGP2013} can
help, but it will be necessary to fit the lattice data for $\P(Q^2)$ to a 
continuous function of $Q^2$ in order to evaluate the integral.
The problem then becomes that of finding a theoretically well-founded
functional form for the $Q^2$ dependence of $\P(Q^2)$, so that this
functional form can be fitted to available data, after which the
integral in Eq.~(\ref{amu}) is performed using the fitted function.

A number of fit functions have been used and/or proposed recently.
One class of fit functions is based on Vector Meson Dominance (VMD)
\cite{AB2007,FJPR2011,BDKZ2011}, another class on Pad\'e Approximants
(PAs) \cite{DJJW2012,ABGP2012}, while a position-space version of VMD-type 
fits was recently proposed in Ref.~\cite{FJMW}.  VMD-type fits, as well as the 
PAs used in Ref.~\cite{DJJW2012} do not represent members of a sequence of 
functions guaranteed to converge to the actual vacuum polarization, whereas
the PAs of Ref.~\cite{ABGP2012} do.   Thus, theoretical prejudice would
lead one to choose the  PAs of Ref.~\cite{ABGP2012} as the appropriate set of 
functions to fit lattice data for the vacuum polarization.

However, this does not guarantee that any particular fit to lattice data
for the vacuum polarization will yield an accurate estimate of $a_\m^{\rm HLO}$
with a reliable error.   This depends not only on the theoretical validity of
the fit function, but also, simply, on the availability of good data.   
Moreover, even if a sequence of PAs converges (on a certain $Q^2$ interval), not much is known in practice about how fast its rate of convergence may be.
For example, if the convergence is very slow given a certain lattice data set, it could be that only 
PAs with a number of parameters far beyond the reach of these data
give a numerically adequate representation of the true vacuum 
polarization, for the goal of computing $a_\m^{\rm HLO}$ to a 
phenomenologically interesting
accuracy.

It would therefore be useful to have a good model, in which the ``exact''
answer is known.   One can then investigate any given fitting method,
and ask questions such as whether a good fit (for instance, as measured
by the $\c^2$ per degree of freedom) leads to an accurate result for
$a_\m^{\rm HLO}$.  If the model is a good model, this will not only
test the theoretical validity of a given fit function, but also how well this fit works, given a required accuracy, and given a set of data for $\P(Q^2)$.  In other words, it will give us a reliable quantitative estimate of the systematic error. 

Such a model is available for the vacuum polarization.   The $I=1$
non-strange hadronic vector spectral function has been very accurately
measured in hadronic $\t$ decays.  From this spectral function, one can, using
a dispersion relation, construct the corresponding component of the
vacuum polarization, if one has a reliable theoretical representation
for the spectral function beyond the $\t$ mass.   Such a representation
was constructed in Refs.~\cite{BCGJMOP2011,BGJMMOP2012} from OPAL data for this
spectral function \cite{OPAL}.   The thus obtained vacuum polarization
is closely related to the $I=1$ component of the vacuum polarization
obtained from $\s(e^+e^-\to\g\to\mbox{hadrons})$.   

Three points are relevant to understanding the use of the term ``model''
for the resulting $I=1$ polarization function, in the context of the
underlying $a_\m^{\rm HLO}$ problem. First, $a_\m^{\rm HLO}$ is related directly
to $\s(e^+ e^- \rightarrow \g \rightarrow \mbox{hadrons})$ \cite{BM1961}
and the
associated electromagnetic (EM) current polarization function, which, unlike
the model, has both an $I=1$ and $I=0$ component. Second, even for the $I=1$
part there are subtleties involved in relating the spectral functions
obtained from $\s(e^+ e^- \rightarrow \g\rightarrow\mbox{hadrons})$
and non-strange $\t$ decays \cite{Detal2009,DM2010}. Finally, since the $\t$ data
extends only up to $s=m_\t^2$, a model representation is required for the
$I=1$ spectral function beyond this point. 

In fact, we consider the pure
$I=1$ nature of the model polarization function an advantage for the purposes
of this study, as it corresponds to a simpler spectral distribution
than that of the EM current (the latter involving also the light quark and
$\bar{s}s$ $I=0$ components). Working with the $\t$ data also allows us to
avoid having to deal with the discrepancies between the determinations of
the $\p^+\p^-$ electroproduction cross-sections obtained by different
experiments \cite{CMD2pipi07,SNDpipi06,BaBarpipi12,KLOEpipi12}.\footnote{Figs.~48 and 50 of
Ref.~\cite{BaBarpipi12} provide a useful overview of the current situation.} We should
add that, though a model is needed for the part of the $I=1$ spectral
function beyond $s=m_\t^2$, for the low $Q^2$ values relevant to $a_\m^{\rm HLO}$,
the vacuum polarization we construct is very insensitive to the
parametrization used in this region.  Finally, we note that the model vacuum
polarization satisfies, by construction, the same analyticity
properties as the real vacuum polarization.
  In particular, the subtracted model vacuum polarization
is equal to $Q^2$ times a Stieltjes function \cite{ABGP2012}.
 We thus expect our model to be an excellent model
for the purpose of this article, which is to test a number of methods
that have been employed in fitting the $Q^2$ dependence of the
vacuum polarization to lattice data, and not to determine
the $I=1$ component of $a_\m^{\rm HLO}$ from $\t$ spectral data.

This article is organized as follows.  In the following two sections, 
we construct the model, and define the fit functions we will consider here.
Throughout this paper, we will consider only VMD-type
fits, which have been extensively used, and PA fits of the type defined
in Ref.~\cite{ABGP2012}.\footnote{For other PA fits considered in the 
literature, we are not aware of any convergence theorems.}
In Sec.~\ref{lattice}, we use the model and a typical covariance matrix
obtained in a lattice computation to generate fake ``lattice'' data sets,
which are then fitted in Sec.~\ref{fits}.   We consider both correlated
and diagonal (``uncorrelated'') fits, where in the latter case errors
are computed by linear propagation of the full 
data covariance matrix through the fit. From these fits, estimates
for $a_\m^{\rm HLO}$ with errors are obtained, and compared with
the exact model value in order to test the accuracy of the fits. 
Section~\ref{conclusion} contains our conclusions.

%%####%%
%\newpage
\section{\label{model} Construction of the model}
%%####%
The non-strange, $I=1$
subtracted vacuum polarization is given by the dispersive integral
\begin{equation}
\label{disp}
\tP(Q^2)=\P(Q^2)-\P(0)=-Q^2\int_{4m_\p^2}^\infty dt\;\frac{\r(t)}{t(t+Q^2)}\ ,
\end{equation}
where $\r(t)$ is the corresponding spectral function, and $m_\p$ the pion 
mass.  In order to construct our model for $\tP(Q^2)$,
we split this integral into two parts:  one with $4m_\p^2\le t\le s_{min}
\le m_\t^2$, and one with $s_{min}\le t<\infty$.  In the first region,
we use OPAL data to estimate the integral by a simple Riemann sum:
\begin{equation}
\label{region1}
\left(\P(Q^2)-\P(0)\right)_{t\le s_{min}}=-Q^2\D t\sum_{i=1}^{N_{min}}\;
\frac{\r(t_i)}{t_i(t_i+Q^2)}\ .
\end{equation}
Here the $t_i$ label the midpoints of the bins from the lowest bin $i=1$
to the highest bin $N_{min}$ below $s_{min}=N_{min}\D t$, and $\D t$ is the bin width,
which for the OPAL data we use is equal to $0.032$~GeV$^2$.  For
the contribution from
the spectral function above $s_{min}$, we use the representation
\begin{subequations}
\begin{eqnarray}
\label{region2}
\left(\P(Q^2)-\P(0)\right)_{t\ge s_{min}}&=&-Q^2\int_{s_{min}}^\infty dt\;\frac{\r_{t\ge s_{min}}(t)}{t(t+Q^2)}\ ,\label{region2a}\\
\r_{t\ge s_{min}}(t)&=&\r_{\rm pert}(t)+e^{-\d-\g t}\sin(\a+\b t)\ ,\label{region2b}
\end{eqnarray}
\end{subequations}
where $\r_{\rm pert}(t)$ is the perturbative part calculated to five loops
in perturbation theory, expressed in terms of $\a_s(m_\t^2)$ \cite{BCK2008},
with $m_\t$ the $\t$ mass.
The oscillatory term is our representation of the duality-violating part,
and models the presence of resonances in the measured spectral
function. This representation of the spectral function was extensively
investigated in Refs.~\cite{BCGJMOP2011,BGJMMOP2012}, and found to give a very good
description of the data between $s_{min}=1.504$~GeV$^2$ and $m_\t^2$.
Figure~\ref{f2} shows the comparison between the data and the
representation~(\ref{region2b}) for this value of $s_{min}$;
the blue continuous curve shows the representation we will be 
employing here. Our central values
for $\a_s(m_\t^2)$, $\a$, $\b$, $\g$ and $\d$ have been taken from the
FOPT $w=1$ finite-energy sum rule fit of Ref.~\cite{BGJMMOP2012}:\footnote{The
final one or two digits of these parameter values are not significant in view
of the errors obtained in Eq.~(5.3) of Ref.~\cite{BGJMMOP2012}, but these 
are the values we used to construct the model.}
\begin{eqnarray}
\label{w1FOPT}
\a_s(m_\t^2)&=&0.3234   \ ,\\
\a&=&-0.4848   \ ,\nonumber\\
\b&=&3.379~\mbox{GeV}^{-2}  \ ,\nonumber\\
\g&=&0.1170~\mbox{GeV}^{-2}   \ ,\nonumber\\
\d&=&4.210   \ .\nonumber
\end{eqnarray}   
The low-$Q^2$ part of the function
$\P(Q^2)$ obtained through this strategy is shown as the blue curve
in the left-hand panel of Fig.~\ref{f1}.

%%%%%%%%%%%%%%%%%%%
\begin{figure}[t]
\centering
\includegraphics[width=4in]{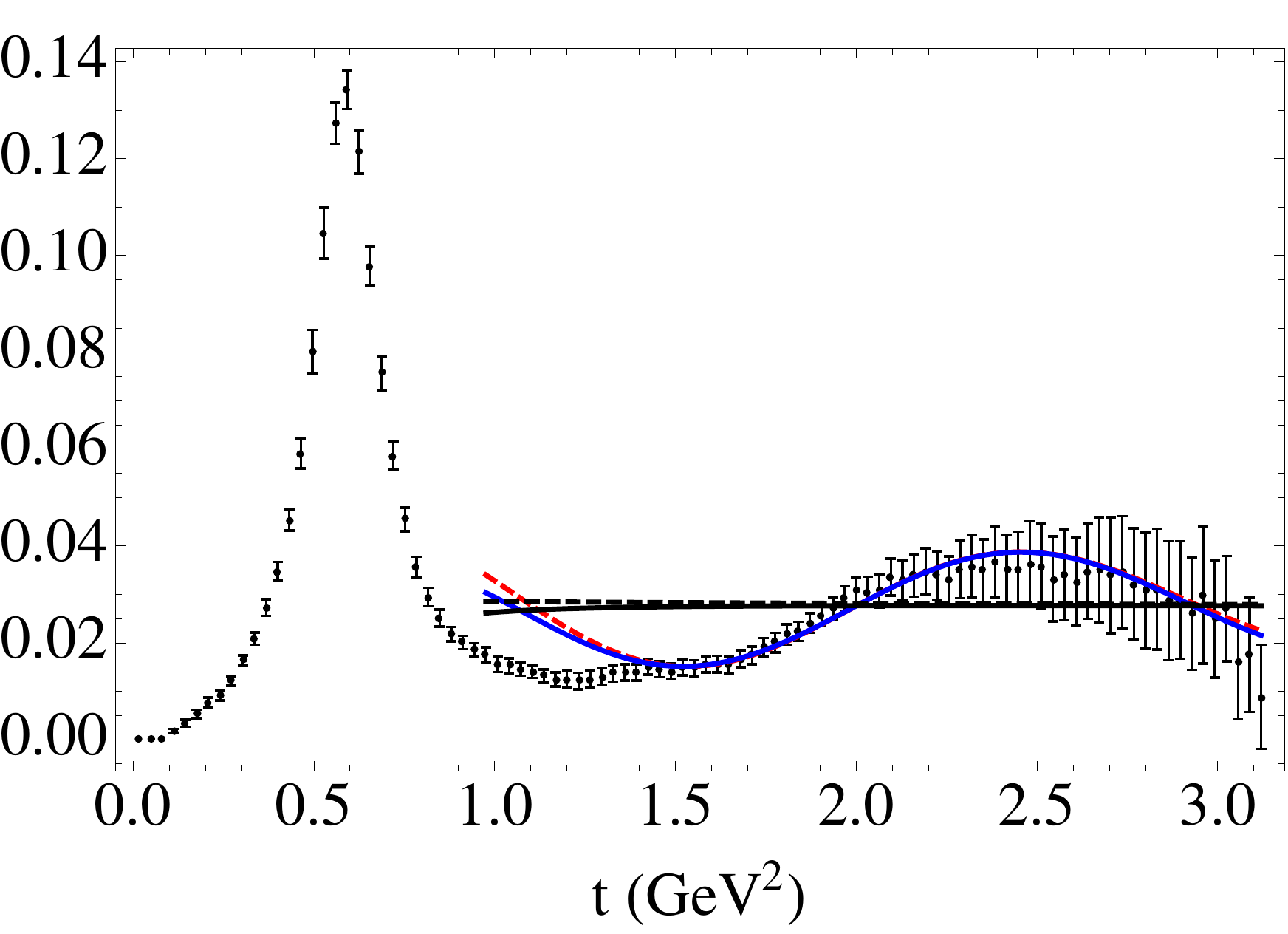}
\floatcaption{f2}{The $I=1$ non-strange vector spectral function, 
from Ref.~\cite{BGJMMOP2012}, as a function of $t$. The data are from 
OPAL \cite{OPAL}, the curves are theoretical representations 
obtained from the $w=1$ finite-energy sum rule discussed 
in Refs.~\cite{BCGJMOP2011,BGJMMOP2012}.}
\vspace*{2ex}
\end{figure}
%%%%%%%%%%%%%%%%%%%

As in Ref.~\cite{ABGP2012} we will take as a benchmark the low- 
and medium-$Q^2$ part of $a_\m^{\rm HLO}$, 
\begin{equation}
\label{amu1}
\ta_\m^{{\rm HLO},Q^2\le 1}=4\a^2\int_0^{1~{\rm GeV}^2} dQ^2\,f(Q^2)\left(\P(0)-\P(Q^2)\right)\ .
\end{equation}
To make it clear that we are computing this quantity from $\tP(Q^2)$
defined from Eqs.~(\ref{disp})-(\ref{w1FOPT}), and not from $\P^{\rm em}(Q^2)$, 
we will use the symbol $\ta_\m$, instead of $a_\m$, in the rest of this 
article.

Using the OPAL data as described above, and fully propagating errors,%
\footnote{Taking into account the OPAL data covariance
matrix, the parameter covariance matrix for the parameters in 
Eq.~(\ref{region2b}), as well as the correlations between OPAL data 
and the parameters.} we find the value
\begin{equation}
\label{amuvalue}
\ta_\m^{{\rm HLO},Q^2\le 1}=1.204(27)\times 10^{-7}\ .
\end{equation}

In our tests of lattice data in Sec.~\ref{fits} below, 
we will declare the model to be ``exact,'' and see how 
various fits to fake lattice data generated from the model will fare in 
reproducing this exact value. For our purposes, it is sufficient to have
a four-digit ``exact'' value, which we take to be
\begin{equation}
\label{amuexact}
\ta_{\m,{\rm model}}^{{\rm HLO},Q^2\le 1}=1.204\times 10^{-7}\ .
\end{equation}

We close this section with a few remarks. In the region 
$0\le Q^2\le 1$~GeV$^2$, the model we constructed
for $\tP(Q^2)$ is very insensitive to both the detailed
quantitative form of Eq.~(\ref{region2b}), 
as well as to the choice of $s_{min}$.
Moreover, the precise quantitative values that we obtain for $\tP(Q^2)$
as a function of $Q^2$ are not important.   What is important is that this is 
a very realistic model, based on hadronic data which are very well 
understood in the framework of QCD, for the $I=1$ part of $\P^{\rm em}(Q^2)$.

%%####%%
%\newpage
\section{\label{functions} Fit functions}
%%####%
We will consider two classes of fit functions to be employed in fits
to data for $\P(Q^2)$. The first class of functions involves PAs of the form
\begin{equation}
\label{PA}
\P(Q^2)=\P(0)-Q^2\left(a_0+\sum_{k=1}^K\frac{a_k}{b_k+Q^2}\right)\ .
\end{equation}
For $a_0=0$, the expression between parentheses is a $[K-1,K]$ Pad\'e;
if also $a_0$ is a parameter, it is a $[K,K]$ Pad\'e.  With $a_{k\ge 1}>0$
and $b_k>b_{k-1}>\dots>b_1>4m_\p^2$, these PAs constitute a 
sequence converging to the exact vacuum polarization in the sense 
described in detail in Ref.~\cite{ABGP2012}. With ``good enough'' data, 
we thus expect that, after fitting the data, one or more of these PAs 
will provide a numerically accurate representation of $\P(Q^2)$ on a compact 
interval for $Q^2$ on the positive real axis. For each such fit,
we may compute $\ta_\m^{{\rm HLO},Q^2\le 1}$, and compare the result
to the exact model value.  Of course, the aim of this article is to 
gain quantitative insight into what it means for the data to be 
``good enough,'' as well as into what order of PA might be required to achieve
a given desired accuracy in the representation of $\P(Q^2)$ at low $Q^2$.

We note that in the model, by construction we have that $\tP(0)=0$.
In contrast, a lattice computation yields only the unsubtracted 
$\P(Q^2)$ at non-zero values of $Q^2$.\footnote{A recent paper proposed 
a method for computing $\P(0)$ directly on the lattice \cite{DPT2012}, 
whereas another recent paper proposed to obtain $\P(Q^2)$ at and near 
$Q^2=0$ by analytic continuation \cite{FHHJPR2013}. Since we do not know yet
what the size of the combined statistical and systematic errors on 
$\P(0)$ determined in such ways will turn out to be, we do not consider 
these options in this article.}
It thus appears that the model does not quite match the lattice framework
it is designed to simulate. However, if in the test fits we treat
$\P(0)$ in Eq.~(\ref{PA}) as a free parameter, we discard the information that
$\tP(0)=0$ in the model, and we can use the fake data 
generated from the model as a test case for the lattice. In other words,
if we treat $\P(0)$ in Eq.~(\ref{PA}) as a free parameter, we can think of the
model vacuum polarization as $\P(Q^2)$ in a
scheme in which $\P(0)$ happens to vanish, rather than as $\tP(Q^2)$.
This turns out to be a very important observation, because even 
if a PA or VMD-type fit does a good job of fitting the overall 
$Q^2$ behavior over a given interval, it is generally
difficult for these fits to yield the correct curvature at very low $Q^2$.
Because the integral in Eq.~(\ref{amu}) is dominated by the low-$Q^2$ region,
this effect can lead to significant deviations of $\ta_\m^{{\rm HLO},Q^2\le 1}$
from the exact model value, as we will see below.

We will also consider VMD-type fits, which have been widely used in the
literature. Typical VMD-type fits have the form of Eq.~(\ref{PA}), but with the
lowest pole, $b_1$, fixed to the $\r$ mass, $b_1=m_\r^2$.   We will
consider two versions: straight VMD, obtained by taking $K=1$
in Eq.~(\ref{PA}) and setting $a_0=0$, and VMD$+$, which is similar but with
$a_0$ a free parameter. Such VMD-type fits have been employed 
previously \cite{DJJW2012,AB2007,FJPR2011,BDKZ2011,BFHJPR2013}. We 
emphasize that VMD-type fits, despite their resemblance to the PAs 
of Eq.~(\ref{PA}), are {\it not} of that type. The exact function $\P(Q^2)$
has a cut at $Q^2=-4m_\p^2$, which has to be reproduced by the
gradual accumulation of poles in Eq.~(\ref{PA}) toward that value. If instead 
we choose the lowest pole at the $\r$ mass, the fit function is a model 
function based on the intuitive picture of vector meson dominance, and 
is definitely not a member of the convergent sequence introduced 
in Ref.~\cite{ABGP2012}. However, as already emphasized in Sec.~\ref{intro}, 
the aim here is to investigate the quality of various fits on test data, 
without theoretical prejudice. We will thus investigate both PA and 
VMD-type fits in the remainder of this article.

In Ref.~\cite{BDKZ2011} also a VMD-type fit with two poles, obtained by 
choosing $K=2$, $a_0=0$ and $b_1=m_\r^2$ has been considered.
In our case, such a fit turns out to not yield any extra information
beyond VMD+: we always find that $b_2$ is very large, and $a_2$ and
$b_2$ are very strongly correlated, with the value of $a_2/b_2$
equal to the value of $a_0$ found in the VMD+ fit. The reason this
does not happen in Ref.~\cite{BDKZ2011} is probably that in that case also the
connected part of the $I=0$ component is included in $\P(Q^2)$, and this
component has a resonance corresponding to the octet component of 
the $\phi$-$\omega$ meson pair. In our case, in which only the $I=1$
component is present, these two-pole VMD-type
fits never yield any information beyond the VMD+ fits.

%%####%%
%\newpage
%\begin{boldmath}
%\section{\label{tau} Fits using $\t$ data}
%\end{boldmath}
%%####%

%%####%%
%\newpage
\section{\label{lattice} The generation of fake lattice data}
%%####%
%%%%%%%%%%%%%%%%%%%
\begin{figure}[t]
\centering
\includegraphics[width=4in]{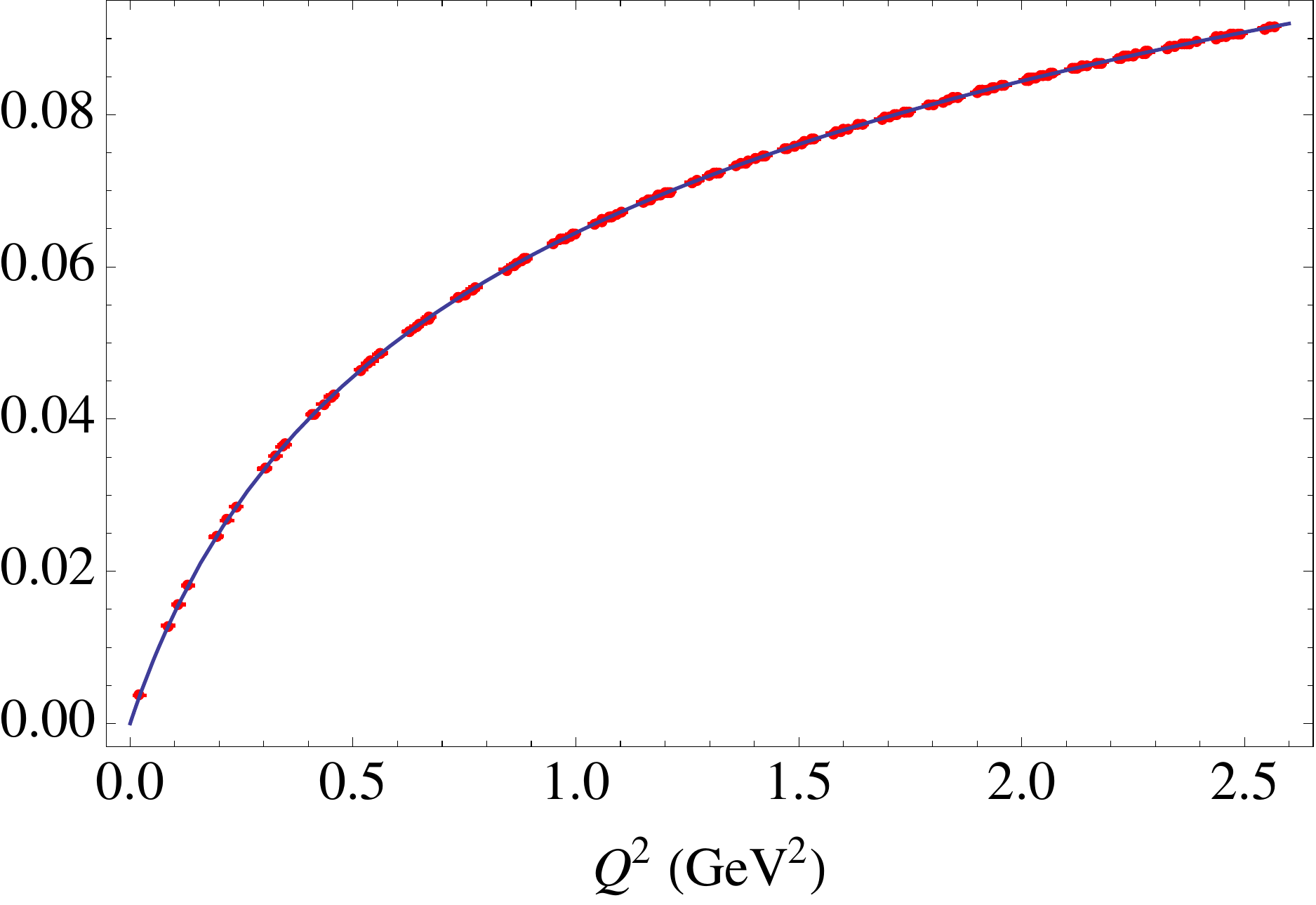}
\floatcaption{f3}{Fake data set constructed in Sec.~\ref{lattice} and the
model for $\P(0)-\P(Q^2)$ constructed in Sec.~\ref{model} (thin blue curve).}
\vspace*{2ex}
\end{figure}
%%%%%%%%%%%%%%%%%%%
In order to carry out the tests, we need data that correspond to a world
described by our model, and that resemble a typical set of lattice data.
In order to construct such a data set, we proceed as follows. First,
we choose a set of $Q^2$ values. The $Q^2$ values we will consider
are those available on an $L^3\times T=64^3\times 144$ lattice 
with periodic boundary conditions, and an inverse lattice spacing 
$1/a=3.3554$~GeV. The smallest
momenta on such a lattice in the temporal and spatial directions are
\begin{eqnarray}
\label{exmom}
Q&=&\left(0,0,0,\frac{2\p}{aT}\right)\quad\rightarrow\quad Q_1^2=0.02143~\mbox{GeV}^2\ ,\\
Q&=&\left(0,0,0,\frac{4\p}{aT}\right)\quad\rightarrow\quad Q_2^2=0.08574~\mbox{GeV}^2\ ,\nonumber\\
Q&=&\left(\frac{2\p}{aL},0,0,0\right)\quad\rightarrow\quad Q_3^2=0.1085~\mbox{GeV}^2\ ,\nonumber
\end{eqnarray}
\etc\  Next, we construct a multivariate Gaussian distribution with central
values $\P(Q_i^2)$, $i=1,2,\dots$, and a typical covariance matrix obtained
in an actual lattice computation of the vacuum polarization on this lattice.
The covariance matrix we employed is the covariance matrix for the
$a=0.06$~fm data set considered in Ref.~\cite{ABGP2012}. The fake data
set is then constructed by drawing a random sample from this 
distribution.\footnote{We used the Mathematica routines 
{\tt MultinormalDistribution} and {\tt RandomVariate}.}
The data points shown in Fig.~\ref{f1} are the first three data points 
of this fake data set. The full data set is shown in Fig.~\ref{f3}. 
We will refer to this as the ``lattice'' data set.

Below, we will also have use for a ``science-fiction'' data set. This second
data set is obtained exactly as the fake data set described above, except
that we first divide the lattice covariance matrix by 10000, which corresponds
to reducing diagonal errors by a factor 100. After this reduction, the data
set is generated as before. We refer to this as the ``science-fiction'' data 
set because it seems unlikely that a realistic lattice data set with such 
good statistics will exist in the near future. However, this second data
set will allow us to gain some additional insights in the context of this 
model study.

%%####%%
%\newpage
\section{\label{fits} Fits to the fake lattice data}
%%####%
\begin{table}[t]
\begin{center}
\begin{tabular}{|c|c|c|c||c|c|c|}
\hline
Fit & $\ta_\m^{{\rm HLO},Q^2\le 1}\times 10^7$  & $\sigma$
& $\c^2$/dof & $\ta_\m^{{\rm HLO},Q^2\le 1}\times 10^7$&$\s$ & $\c^2$/dof\\
\hline
PA $[0,1]$   &   0.8703(95)    &  &     285/46 & 0.6805(45) && 1627/84\\
PA $[1,1]$   &  1.116(22)    & 4  &      61.4/45 & 1.016(12) & 16 & 189/83\\
PA $[1,2]$   &  1.182(43)    & 0.5  &     55.0/44 & 1.117(22) & 4 & 129/82\\
PA $[2,2]$    &  1.177(58)      & 0.5 &      54.6/43  & 1.136(38) & 1.8 & 128/81\\
\hline
VMD & 1.3201(52)  &  & 2189/47 & 1.3873(44) & & 18094/85\\
VMD+ & 1.0658(76)  & 18 & 67.4/46 & 1.1041(48) & 21& 243/84\\
\hline
\end{tabular}
\end{center}
\begin{quotation}
\floatcaption{table1}{Various correlated fits of the ``lattice'' data set 
constructed in Sec.~\ref{lattice} on the interval $0<Q^2\le 1$ GeV$^2$ 
(left of the vertical double line), or on the interval 
$0<Q^2\le 1.5$ GeV$^2$ (right of the vertical double line). 
For a more detailed description, see the text.}
\end{quotation}
\vspace*{-4ex}
\end{table}%

In this section, we will present and discuss the results of a number of fits, 
based on the data sets constructed in Sec.~\ref{lattice}.

%%####%%
%\newpage
\subsection{\label{fake1} ``Lattice'' data set}
%%####%
Table~\ref{table1} shows the results of a number of correlated fits of the lattice data
set to the functional forms defined in Sec.~\ref{functions}.  To the left of the
vertical double line the fitted data are those in the interval
$0<Q^2\le 1$ GeV$^2$; to the right the fitted data are those in the interval
$0<Q^2\le 1.5$ GeV$^2$. In each of these two halves,
the left-most column shows the fit function, and
the second column gives the value of $\ta_\m^{{\rm HLO},Q^2\le 1}$
obtained from the fit, with the $\c^2$ fit error between parentheses.
The ``pull'' $\s$ in the third column is defined as
\begin{equation}
\label{pull}
\s=\frac{|\mbox{exact\ value}-\mbox{fit\ value}|}{\mbox{error}}\ .
\end{equation}
For instance, with the exact value of Eq.~(\ref{amuexact}), we have for the
$[1,1]$ PA on the interval $0<Q^2\le 1$ GeV$^2$
that $\s=|1.204-1.116|/0.022=4$.   The fourth column gives the $\c^2$
value per degree of freedom (dof) of the fit.

Of course, the pull can only be computed because we know the exact
model value. This is precisely the merit of this model study: it gives
us insight into the quality of the fit independent of the $\c^2$ value. 
Clearly, the fit does a good job if the pull is of order one, because if 
that is the case, the fit error covers the difference between the exact 
value and the fitted value.  

The primary measure of the quality of the fit is the value of $\c^2/$dof.   
This value clearly rules out $[0,1]$ PA and VMD as good fits. In this
case, we do not even consider the pull: these functional forms clearly
just do not represent the data very well. However, in all other cases,
one might consider the value of $\c^2/$dof to be reasonable, although
less so for fits on the interval $0<Q^2\le 1.5$ GeV$^2$. However,
only the $[1,2]$ and $[2,2]$ PAs have a good value for the pull
for fits on the interval $0<Q^2\le 1$ GeV$^2$, whereas
the pull for the $[1,1]$ PA and VMD$+$ is bad: the fit error does not
nearly cover the difference between the true (\ie, exact model) value 
and the fitted value.   Note that with the errors of the ``lattice''
data set even the best result, from the $[2,2]$ PA, only
reaches an accuracy of 5\% for $\ta_\m$.  
On the interval $0<Q^2\le 1.5$ GeV$^2$ all fits 
get worse, both as measured by $\c^2$ and $\s$,
and only the $[2,2]$ PA may be considered acceptable.

%%%%%%%%%%%%%%%%%%%
\begin{figure}[t]
\centering
\includegraphics[width=2.9in]{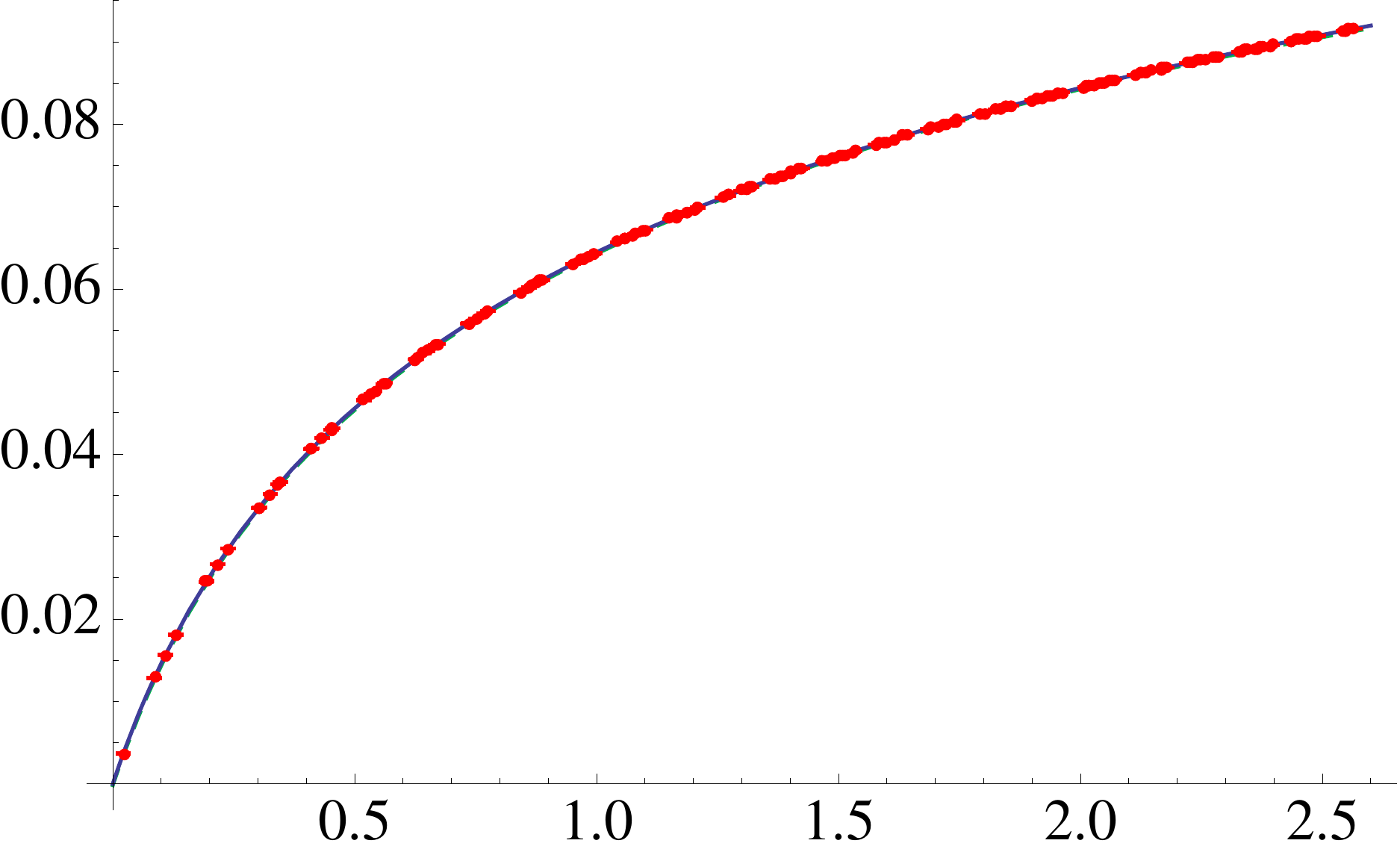}
\hspace{.1cm}
\includegraphics[width=2.9in]{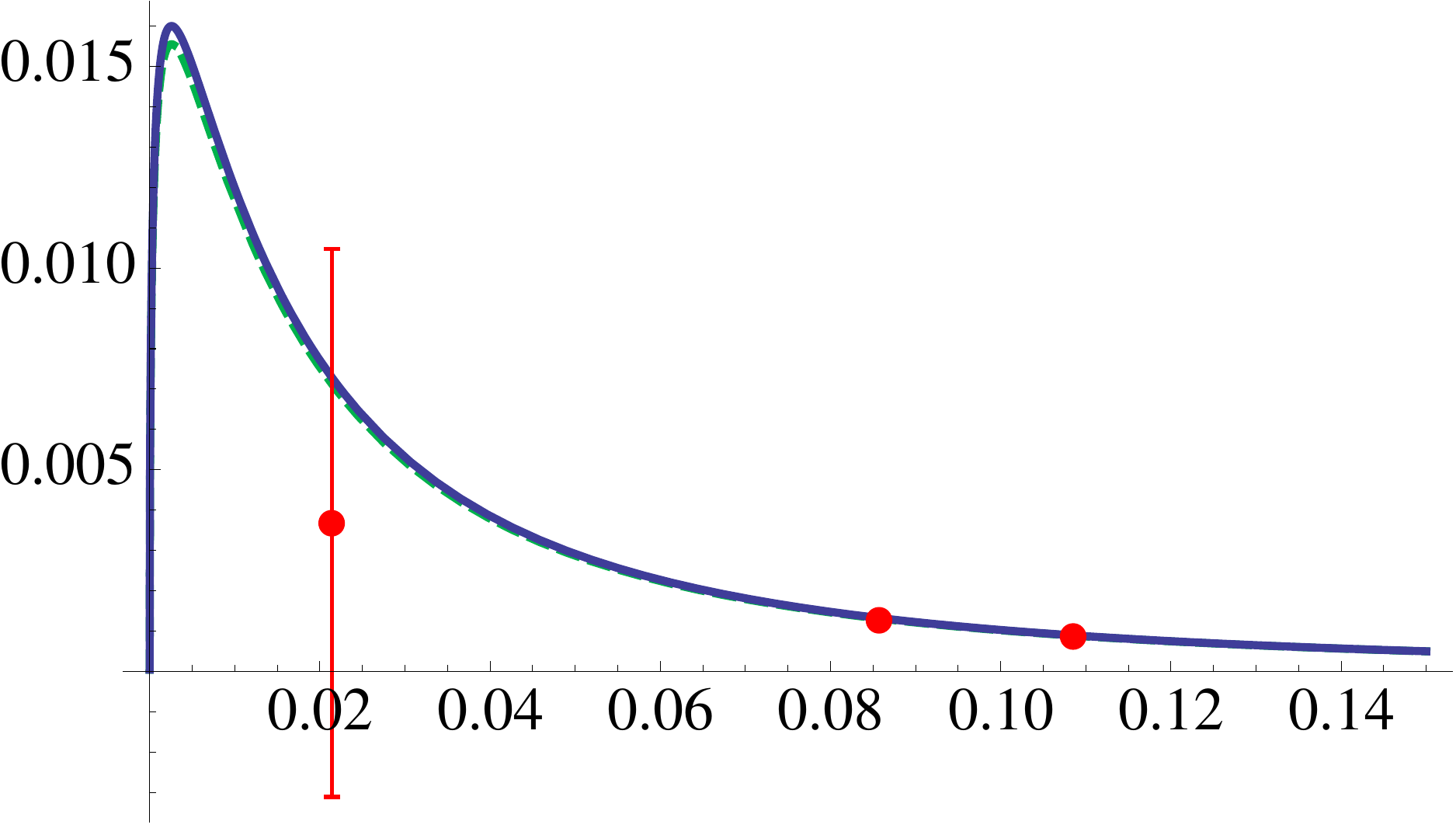}
\floatcaption{PA12}{The $[1,2]$ PA fit on the interval
$0<Q^2\le 1$~GeV$^2$ of Table~\ref{table1}: green dashed curves,
in comparison with the model (blue solid curves). The left panel
shows the vacuum polarization and the right panel shows the blown-up
low-$Q^2$ region of the integrand of Eq.~(\ref{amu}); red points are
``lattice'' data points. Axes and units as in Fig.~\ref{f1}.}
\vspace*{2ex}
\end{figure}
%%%%%%%%%%%%%%%%%%%

%%%%%%%%%%%%%%%%%%%
\begin{figure}[t]
\centering
\includegraphics[width=2.9in]{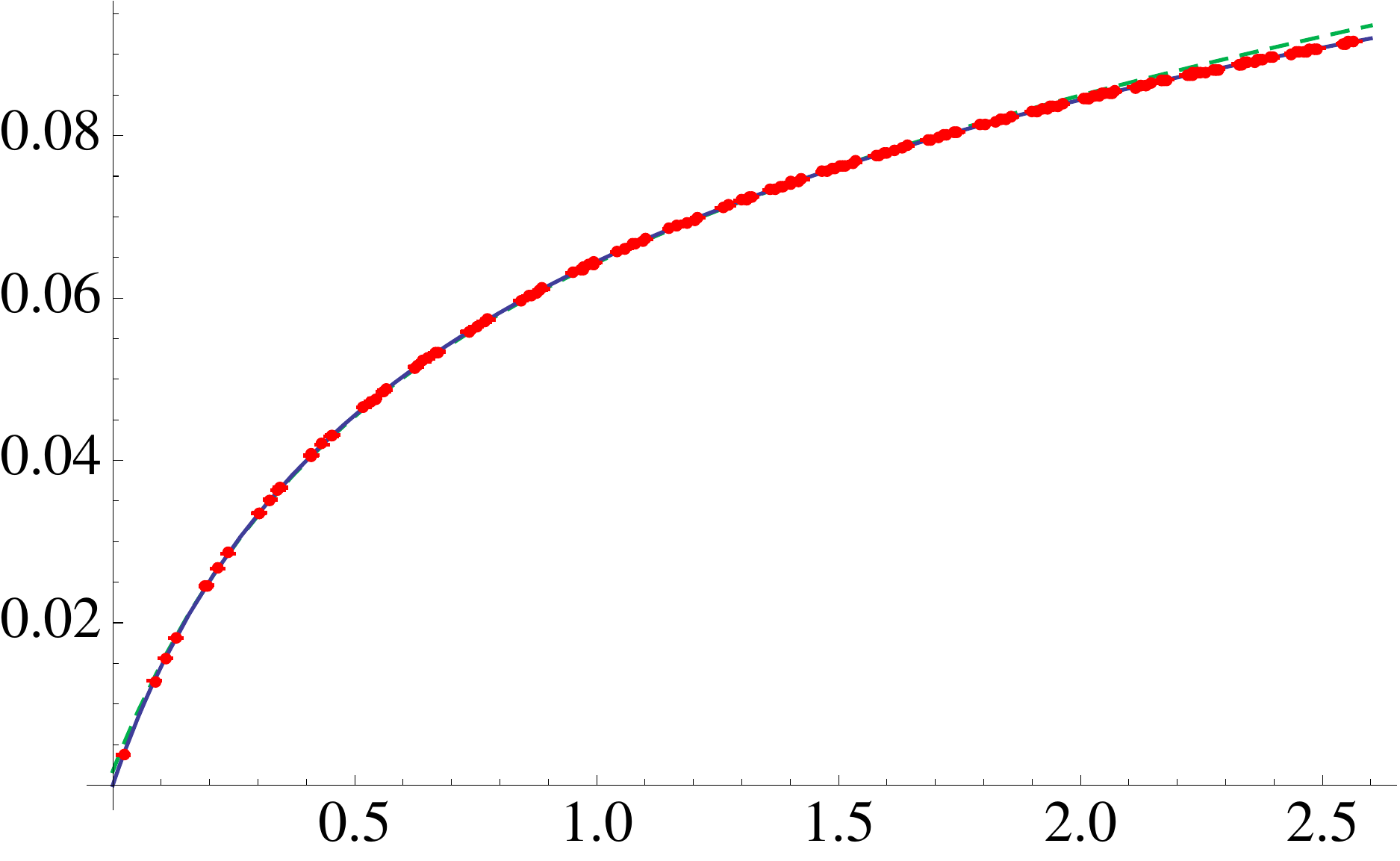}
\hspace{.1cm}
\includegraphics[width=2.9in]{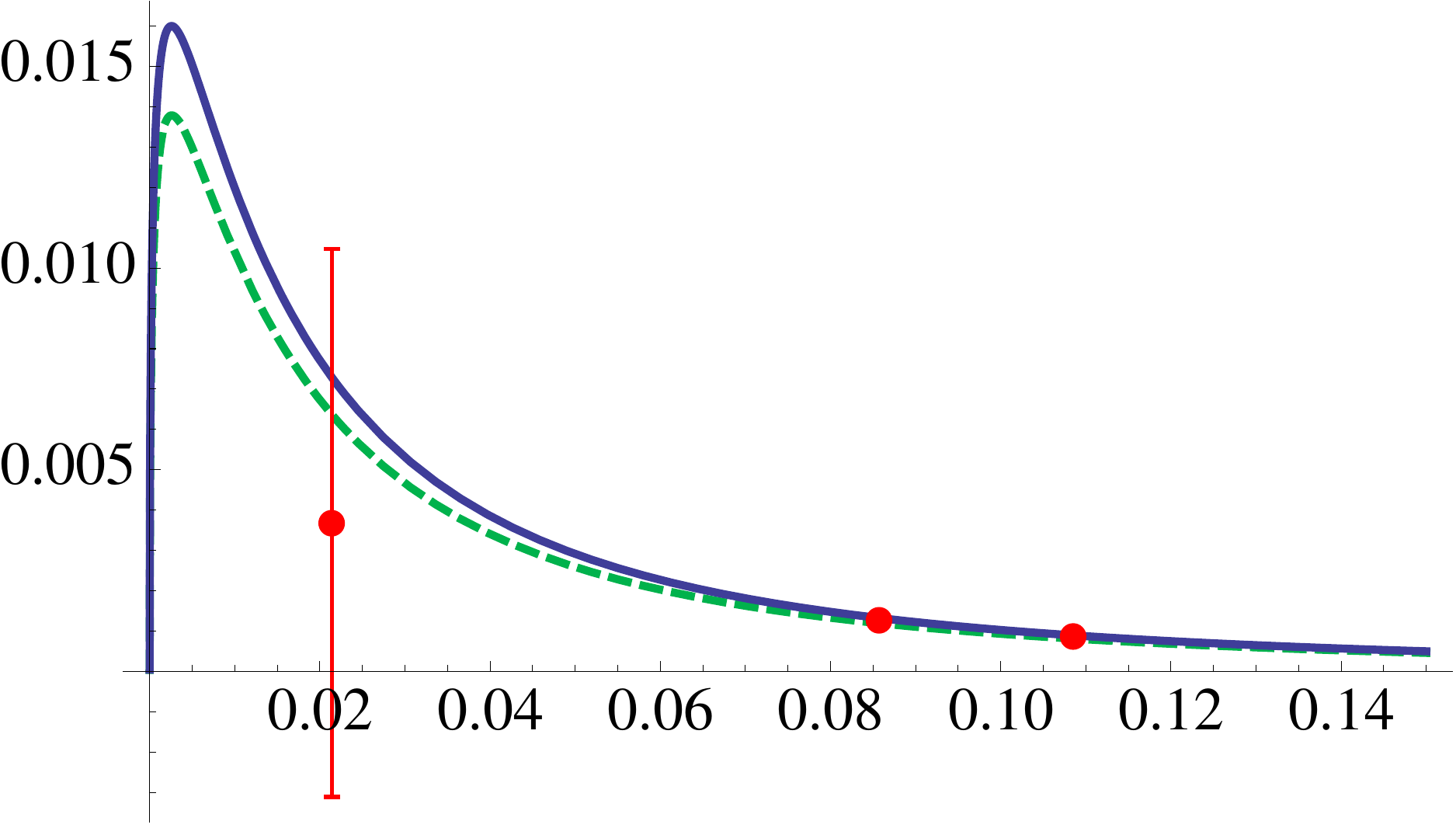}
\floatcaption{VMDplus}{The VMD$+$ fit on the interval
$0<Q^2\le 1$~GeV$^2$ of Table~\ref{table1}: green dashed curves,
in comparison with the model (blue solid curves). The left panel
shows the vacuum polarization and the right panel shows the blown-up
low-$Q^2$ region of the integrand of Eq.~(\ref{amu}); red points are
``lattice'' data points. Axes and units as in Fig.~\ref{f1}.}
\vspace*{2ex}
\end{figure}
%%%%%%%%%%%%%%%%%%%

For illustration, we show the $[1,2]$ PA and VMD$+$ fits on the 
interval $0<Q^2\le 1$ GeV$^2$ in Figs.~\ref{PA12} and \ref{VMDplus}. 
The left-hand panels show the fit over a wider range of $Q^2$, including 
the full set of $Q^2$ values employed in the fit, while the right-hand 
panels focus on the low $Q^2$ region of the integrand in Eq.~(\ref{amu}) of 
primary relevance to $\ta_\m^{{\rm HLO},Q^2\le 1}$, which contains only a few of the
$Q^2$ fit points. The blue solid curve shows the exact model, the green
dashed curve the fit, and the red points are the lattice data. Both fits
to the vacuum polarization look like good fits (confirmed
by the $\c^2/$dof values) when viewed from the perspective of the left-hand
panels. A clear distinction, however, emerges between the
$[1,2]$ PA and VMD$+$ cases when one focuses on the low-$Q^2$
region shown in the right-hand panels. In these panels, the PA fit
follows the exact curve very closely, while the VMD$+$ fit undershoots the
exact curve by a significant amount, as quantified by the pull.
Looking at the left hand panels in Figs.~\ref{PA12} and \ref{VMDplus}, 
one would never suspect the difference in the results for $\ta_\m^{{\rm HLO},Q^2\le 1}$ 
illustrated in  the corresponding right hand panels.

%%####%%
%\newpage
\subsection{\label{fake10000} ``Science-fiction'' data set}
%%####%
In Table~\ref{table2}, we show the same type of fits as in Table~\ref{table1},
but now using the ``science-fiction'' data set defined in Sec.~\ref{lattice}. 
The corresponding figures are very similar to Figs.~\ref{PA12} and \ref{VMDplus}, and hence are not shown here.

\begin{table}[t]
\begin{center}
\begin{tabular}{|c|c|c|c|}
\hline
Fit & $\ta_\m^{{\rm HLO},Q^2\le 1}\times 10^7$  & $\s$
& $\c^2$/dof  \\
\hline
PA $[0,1]$   &   0.87782(9)    &  &    1926084/46   \\
PA $[1,1]$   &  1.0991(2)    &  &      51431/45   \\
PA $[1,2]$   &  1.1623(4)     &  &     1340/44 \\
PA $[2,2]$   &  1.1862(15)     & 12  &     76.4/43 \\
PA $[2,3]$    &  1.1965(28)     & 2 &      42.0/42  \\
\hline
VMD & 1.31861(5)  &  & 20157120/47  \\
VMD+ & 1.07117(8)  &  & 70770/46 \\
\hline
\end{tabular}
\end{center}
\begin{quotation}
\floatcaption{table2}{Fits analogous to those reported in Table~\ref{table1}, 
obtained using the ``science-fiction'' data set, for which the covariance 
matrix was reduced by a factor 10000. Fitting interval $0<Q^2\le 1$ GeV$^2$.}
\end{quotation}
\vspace*{-4ex}
\end{table}%

This data set is, of course, quite unrealistic: real lattice data with 
such precision will not soon be generated. But these fits address the 
question of which of the fit functions considered might still be acceptable 
in this hypothetical world, and whether simply decreasing the errors,
in this case by the large factor of 100, rather than also filling in low 
$Q^2$ values, will be sufficient to achieve the goal of 
getting to the desired $\sim 1\%$ accuracy in the determination
of $\ta_\m^{{\rm HLO},Q^2\le 1}$. The answer is barely.

First, we see that the VMD-type fits are completely ruled out already
by the $\c^2$ values. The higher precision data are also more punishing
on the PA fits. By $\c^2$ values, the first three PAs are excluded, in
contrast to Table~\ref{table1}, where only the $[0,1]$ PA is really excluded
by its $\c^2$ value. The $[2,2]$ PA has a possibly reasonable $\c^2$ value,
but its accuracy does not match its precision, with a pull equal to 
12.\footnote{We define the ``precision'' as the error we obtain, while the
``accuracy'' is the difference between the exact and fitted values.}
The more precise data make it possible to perform a $[2,3]$ PA fit, and
this fit is borderline acceptable, given the value of the pull.   

The best fit for each data set yields $a_\m^{{\rm HLO},Q^2\le 1}$ with an
error of 5\% for the lattice data set, down to 0.2\% for the science-fiction
data set. While this means that (real) lattice data with a precision somewhere
in between would yield an error of order 1\% or below, we also see from this
example that that precision does not necessarily translate into an equal
accuracy. We conjecture that in order to increase accuracy, data at
more low-$Q^2$ values than present in the fake data sets considered here
will be needed. While precision data in the region of the peak of the
integrand would be ideal, we suspect that filling in the region between the
two lowest $Q^2$ values in this data set might already be of significant
help.  

%%####%%
%\newpage
\subsection{\label{diagonal} Diagonal fits}
%%####%
It is important to emphasize that the data sets considered here are
constructed such that by definition the covariance matrix employed
is the true covariance matrix, and not some estimator for the true 
one. However, it is possible that for some unknown
reason the covariance matrix we employed for generating the fake
data set is less realistic, even though we took it to come from an actual
lattice computation. For instance, the vacuum polarization of this lattice 
computation contains both $I=1$ and (the connected part of the) $I=0$ 
components, whereas the vacuum polarization considered here has only 
an $I=1$ component.

For this reason,
we also considered diagonal fits, in which instead of minimizing the
$\c^2$ function, we minimize the quadratic form $\cq^2$ obtained
by keeping only the diagonal of the covariance matrix.   However,
our errors take into account the full data covariance matrix by linear 
error propagation.
(For a detailed description of the procedure, see the appendix 
of  Ref.~\cite{BCGJMOP2011}.\footnote{We prefer to refer to this type
of fit as a ``diagonal'' fit, instead of an ``uncorrelated'' fit, as 
the latter phrase suggests, incorrectly, that the off-diagonal part of 
the covariance matrix is completely omitted from the analysis.})

Results of diagonal fits are shown in Tables~\ref{table3} and \ref{table4}.
These tables show fits analogous to those shown in Tables~\ref{table1} 
and \ref{table2}, but instead of taking the full covariance matrix into account
through a $\c^2$ fit, it is only taken into account in the error propagation,
after the fit parameters have been determined from a diagonal fit.

\begin{table}[t]
\begin{center}
\begin{tabular}{|c|c|c|c||c|c|c|}
\hline
& $\ta_\m^{{\rm HLO},Q^2\le 1}\times 10^7$  & $\s$ & $\cq^2$ &$\ta_\m^{{\rm HLO},Q^2\le 1}\times 10^7$ &$\s$ &$\cq^2$\\
\hline
PA $[0,1]$   &  0.997(23)    & 19  & 20.1 & 0.906(15) & 20 & 62.4\\
PA $[1,1]$   & 1.173(74)     & 0.4 &    13.8 & 1.108(39) & 2.5 & 30.3\\
PA $[1,2]$   &  1.30(32)    & 0.3  &   13.55 &1.22(15) & 0.1 & 29.5\\
\hline
VMD & 1.2122(82)  & 1 & 75.2 & 1.2895(69) & 12 & 510\\
VMD+ & 1.083(17) &  7 & 15.0 &1.081(12) & 10 & 30.7\\
\hline
\end{tabular}
\end{center}
\begin{quotation}
\floatcaption{table3}{Fits like those reported in Table~\ref{table1}, but 
using a diagonal fit quality $\cq^2$, and linear propagation of errors.
Fitting interval $0<Q^2\le 1$ GeV$^2$ (left of the vertical
double line), or  $0<Q^2\le 1.5$ GeV$^2$ (right of the vertical
double line).}
\end{quotation}
\vspace*{-4ex}
\end{table}%
The results of these diagonal fits are consistent with, and confirm, the
conclusions one draws from the correlated fits shown in Tables~\ref{table1} 
and \ref{table2}. For PA fits, the only differences are that errors from
the diagonal fits are larger, and the maximum order of the PA for which
we can find a stable fit is one notch lower. Since the fit quality $\cq^2$
is not a $\c^2$ function, its absolute value (per degree of freedom)
has no quantitative probabilistic meaning. But clearly the $[0,1]$,
$[1,1]$, VMD and VMD$+$ fits shown in Table~\ref{table4} are bad
fits, as judged from their $\cq^2$ values. We therefore did not compute
the pull for these fits.   For all other fits in 
Tables~\ref{table3} and \ref{table4} the pull is shown, and consistent 
with that shown in Tables~\ref{table1} and \ref{table2} for PAs of one 
higher order.

Also from these diagonal fits we conclude that the VMD-type fits 
considered here do not work. Amusingly, VMD appears to get it right,
if one takes the VMD fit on the interval $0<Q^2\le 1$~GeV$^2$
in Table~\ref{table3} at face value. However,
this should be considered an accident. If one adds a parameter to
move to a VMD$+$ fit, the value of $\cq^2$ decreases significantly,
as it should, but the pull increases dramatically, showing that VMD$+$ 
is not a reliable fit. This should not happen if the VMD result were
to be reliable itself. Likewise, if we change the fitting interval from
$0<Q^2\le 1$~GeV$^2$ to $0<Q^2\le 1.5$~GeV$^2$, the pull 
increases much more dramatically than for the PA fits.
In addition, both VMD-type fits in Table~\ref{table4}
are bad fits, as judged by the $\cq^2$ values, even though, because
of the same accident, the VMD value for  $a_\m^{{\rm HLO},Q^2\le 1}$
looks very good. Note however, that again the error is nowhere
near realistic as well: we did not compute the pull because of the
large $\cq^2$ value, but its value given the numbers reported is
very large.

We conclude from this example that in order to gauge the reliability of
a fit, ideally one should consider a sequence of fit functions in 
which parameters are systematically added to the fit function. 
This allows one to test the stability of such a sequence of fits, and 
avoid mistakenly interpreting an accidental agreement with the model 
result as an indication that a particular fit strategy is reliable 
when it is not, as happens here for the VMD fit and the specific 
$0$ to $1$ GeV$^2$ fitting window. The PA approach provides a systematic
sequence of fit functions in this respect.

\begin{table}[t]
\begin{center}
\begin{tabular}{|c|c|c|c|c|c|}
\hline
& $\ta_\m^{{\rm HLO},Q^2\le 1}\times 10^7$  & $\s$ & $\cq^2$  \\
\hline
PA $[0,1]$   &  0.99623(23)    &   & 40350  \\
PA $[1,1]$   & 1.12875(68)     &  &    623  \\
PA $[1,2]$   &  1.1762(21)   & 13  &   31.3   \\
PA $[2,2]$    & 1.1904(54)     & 2.5  &  22.1    \\
\hline
VMD & 1.21076(8)  &  & 589751  \\
VMD+ & 1.08341(16)  &   & 4081 \\
\hline
\end{tabular}
\end{center}
\begin{quotation}
\floatcaption{table4}{Diagonal fits like those reported in 
Table~\ref{table3}, but using the ``science-fiction'' data set.
Fitting interval $0<Q^2\le 1$ GeV$^2$.}
\end{quotation}
\vspace*{-4ex}
\end{table}%

%%####%%
%\newpage
\begin{boldmath}
\subsection{\label{medium} The region $1\le Q^2\le 2$~GeV$^2$}
\end{boldmath}
%%####%
While higher-order PAs appear to work reasonably well, in the sense that
their accuracy matches their precision, we also noted that on our fake
lattice data set this is less true when one increases the fit interval
from $0<Q^2\le 1$~GeV$^2$ to  $0<Q^2\le 1.5$~GeV$^2$. At the same
time, one expects QCD perturbation theory only to be reliable above
approximately $2$~GeV$^2$. This leads to the question whether one
can do better on the interval between 0 and 2~GeV$^2$.

As we saw in Sec.~\ref{fits}, the accuracy of the contribution to
$\ta_\m^{{\rm HLO},Q^2\le 1}$ is limited to about 5\% on the lattice
data set, because of the relatively sparse data at low values of $Q^2$. 
We will therefore limit ourselves here to 
a few exploratory comments, in anticipation of future data sets with smaller
errors in the low-$Q^2$ region, and a denser set of $Q^2$ 
values.\footnote{A denser set can be obtained by going to larger volumes,
and/or the use of twisted boundary conditions \cite{DJJW2012,ABGP2013}.}

A possible strategy is to fit the data using a higher-order PA on the interval
$0<Q^2\le Q^2_{\max}$, while computing the contribution between
$Q^2_{\max}$ and 2~GeV$^2$,  $\ta_\m^{{\rm HLO},Q^2_{\max}\le Q^2\le 2}$, 
directly from the data, for some value of $Q^2_{\max}$ such that the PA 
fits lead to reliable results for $\ta_\m^{\rm HLO}$ on the interval 
between 0 and $Q^2_{\max}$. This is best explained by an example, in 
which we choose $Q^2_{max}\approx 1$~GeV$^2$.

The $Q^2$ value closest to 1~GeV$^2$ is $Q^2_{49}=0.995985$~GeV$^2$;
that closest to 2~GeV$^2$ is $Q^2_{129}=2.00909$~GeV$^2$. {}From
our fake data set, using the covariance matrix with which it was generated,
we use the trapezoidal rule to find an estimate
\begin{eqnarray}
\label{trapest}
\ta_\m^{{\rm HLO},Q^2_{49}\le Q^2\le Q^2_{129}}
&=&\half\sum_{i=49}^{128}\left(Q^2_{i+1}-Q^2_i\right)\Bigl(f(Q^2_i)(\P(0)-\P(Q^2_i))\\
&&\hspace{3.5cm}+f(Q^2_{i+1})(\P(0)-\P(Q^2_{i+1}))\Bigr)\nonumber\\
&=&6.925(26)\times 10^{-10}\qquad\mbox{(estimate)}\ .\nonumber
\end{eqnarray}
This is in good agreement with the exact value
\begin{equation}
\label{exactvalue}
\ta_\m^{{\rm HLO},0.995985\le Q^2\le 2.00909}=6.922\times 10^{-10}\qquad
\mbox{(exact)}\ .
\end{equation}
On this interval no extrapolation in $Q^2$ is needed, nor does
the function $f(Q^2)$ play a ``magnifying'' role, so we expect the error
in Eq.~(\ref{trapest}) to be reliable, and we see that this is indeed the case.
In contrast, it is obvious from Fig.~\ref{f1} that estimating 
$\ta_\m^{{\rm HLO},Q^2\le 1}$ in this way would not work. One may
now combine the estimate~(\ref{trapest}) with, for instance, the result
from a fit to the $[1,2]$ PA on the interval $0<Q^2\le 0.995985$~GeV$^2$ 
in order to estimate $\ta_\m^{{\rm HLO},Q^2\le 2.00909}$.\footnote{The 
result from this fit is identical to that on the interval
$0<Q^2\le 1$~GeV$^2$ given in Table~\ref{table1} to the precision
shown in that table.}    The error on this estimate would be determined completely
by that on $\ta_\m^{{\rm HLO},Q^2\le 1}$ coming from the fit, since the
error in Eq.~(\ref{trapest}) is tiny.   Of course, in a complete analysis of this
type, correlations between the ``fit'' and ``data'' parts of $\ta_\m^{\rm HLO}$
should be taken into account, because the values obtained for the fit
parameters in Eq.~(\ref{PA}) will be correlated with the data.   However,
we do not expect this to change the basic observation of this subsection:
The contribution to $\ta_\m^{\rm HLO}$ from the $Q^2$ region between
1 and 2~GeV$^2$ can be estimated directly from the data with a 
negligible error, simply because this contribution to $\ta_\m^{\rm HLO}$
is itself very small (less than 0.6\%).   With better data, this strategy 
can be optimized by varying the value of $Q^2_{max}$. 

\vskip0.8cm
%%####%%
%\newpage
\section{\label{conclusion} Conclusion}
%%####%
In order to compute the lowest-order hadronic vacuum polarization 
contribution $a_\m^{\rm HLO}$ to
the muon anomalous magnetic moment, it is necessary to extrapolate lattice data for the
hadronic vacuum polarization $\P(Q^2)$ to low $Q^2$.   Because of the sensitivity
of $a_\m^{\rm HLO}$ to $\P(Q^2)$ in the $Q^2$ region around $m_\m^2$, one expects a strong
dependence on the functional form used in order to fit data for $\P(Q^2)$ as a function of $Q^2$.   

It is therefore important to test various possible forms of the fit function,
and a good way to do this is to use a model.   Given a model, given a 
set of values of $Q^2$ at which lattice data are available, and given a
covariance matrix typical of the lattice data, one can generate fake
data sets, and test fitting methods by comparing the difference between
the fitted and model values for $a_\m^{\rm HLO}$ with the error on the fitted
value obtained from the fit.   In this article, we carried out such tests,
using a model constructed from the OPAL data for the $I=1$ hadronic
spectral function as measured in $\t$ decays, considering both fit functions
based on Vector Meson Dominance as well as a sequence of Pad\'e
approximants introduced in Ref.~\cite{ABGP2012}.   We took our $Q^2$ 
values and covariance matrix from a recent lattice data set with
lattice spacing $0.06$~fm and volume $64^3\times 144$ \cite{ABGP2012}.

For a fake data set generated for these $Q^2$ values with the given
covariance matrix, we found that indeed it can happen that the precision
of $\ta_\m^{\rm HLO}$ (the analog of $a_\m^{\rm HLO}$ for our model),
\ie, the error obtained from the fit, is much smaller than the accuracy,
\ie, the difference between fitted and exact values.   We considered
correlated fits as well as diagonal fits, and  we also considered
fits to a ``science-fiction'' data set generated with the same covariance matrix scaled
by a factor $1/10000$.

{}From these tests, we conclude that fits based on the VMD-type
fit functions we considered cannot be trusted.   In nearly all cases,
the accuracy is much worse than the precision, and there is no improvement
with the more precise data set with the rescaled covariance matrix.
Adding parameters (VMD$+$) does not appear to help.   Based on
our tests, we therefore call into question the use of VMD-type fits
for the accurate computation of $a_\m^{\rm HLO}$.\footnote{This
includes the recent work in Ref.~\cite{BFHJPR2013}, in which the error on 
$a_\m^{\rm HLO}$ is obtained from a VMD$+$ fit, and in which, reportedly,
the error from PA-type fits is much larger.   
Based on the results we have obtained, we strongly suspect that the error on $a_\m^{\rm HLO}$ in Ref.~\cite{BFHJPR2013} is significantly underestimated.
For example, the results from the $[1,2]$ PA and VMD$+$ fits 
on the interval $0<Q^2\le 1.5$~GeV$^2$ in Table~\ref{table1} are compatible within errors, with the PA error 5 times larger than the VMD$+$ error.  Moreover, in both cases the fit error is too small.}

The sequence of PAs considered here performs better, if one goes to
high enough order.   The order needed may be higher if one uses
more precise data, as shown in the comparison between Tables~\ref{table1}
and \ref{table2}.  Still, with the lattice $Q^2$ values and covariance
matrix of Ref.~\cite{ABGP2012}, the maximum accuracy obtained is of
order a few percent, but at least this is reflected in the errors obtained from the
fits.   Of course, given a certain data set, one cannot add too many 
parameters to the fit, and indeed we find that adding parameters
beyond the $[2,2]$ PA ($[2,3]$ PA for the science-fiction data set)
does not help:  parameters for the added poles at larger $Q^2$ have such
large fitting errors that they do not add any information.   We also
found that PA fits do less well when one increases the fitting interval,
and proposed that the contribution to $a_\m^{\rm HLO}$ from the
region between around 1~GeV$^2$ and the value where QCD
perturbation theory becomes reliable can, instead, be accurately computed
using (for instance) the trapezoidal rule (\seef\ Sec.~\ref{medium}).

We believe that tests such as that proposed in this article should be carried
out for all high-precision computations of $a_\m^{\rm HLO}$. We have
clearly demonstrated that a good $\c^2$ value may {\it not} be 
sufficient to conclude that a given fit is good enough to compute
$a_\m^{\rm HLO}$ with a reliable error. The reason is the ``magnifying effect''
produced by the multiplication of the subtracted vacuum polarization
by the kinematic weight in the integral yielding $a_\m^{\rm HLO}$.
While other useful models (for instance, based on 
$\s(e^+e^-\to\mbox{hadrons})$ data) may also be constructed, 
the model considered here, for the $I=1$ polarization function $\P(Q^2)$,
is already available, and data on this model will be provided
on request. 

\vspace{3ex}
%\newpage
\noindent {\bf Acknowledgments}
\vspace{3ex}

We would like to thank Christopher Aubin and Tom Blum for discussions.
KM thanks the
Department of Physics at the Universitat Aut\`onoma de Barcelona for hospitality.
This work was supported in part by the US Department of Energy, the Spanish Ministerio de Educaci\'on, Cultura y Deporte, under program SAB2011-0074 (MG), the Natural Sciences and
Engineering Research Council of Canada (KM), and by
CICYTFEDER-FPA2011-25948, SGR2009-894,
the Spanish Consolider-Ingenio 2010 Program
CPAN (CSD2007-00042) (SP).

%%%%%%%%%%%%%%%%%%%%%%%%%%%%%%%%%%%%%%%%%%%%%%%%%%%%%%%%%%%%%%%%%%%%%%%%%
%\newpage

\end{document}